# Near-infrared and Optical Nebular-phase Spectra of Type Ia Supernovae SN 2013aa and SN 2017cbv in NGC 5643

Sahana Kumar[1], Eric Y. Hsiao[1], C. Ashall[2], M. M. Phillips[3], N. Morrell[3], P. Hoeflich[1], C. R. Burns[4], L. Galbany[5,6], E. Baron[7], C. Contreras[3], S. Davis[1,8], T. Diamond[9], F. Förster[10,11], M. L. Graham[12], E. Karamehmetoglu[13], R. P. Kirshner[14,15], B. Koribalski[16], K. Krisciunas[17], J. Lu[1], G. H. Marion[18], P. J. Pessi[19], A. L. Piro[4], M. Shahbandeh[1], M. D. Stritzinger[13], N. B. Suntzeff[17], and S. A. Uddin[4]

[1] Department of Physics, Florida State University, 77 Chieftan Way, Tallahassee, FL 32306, USA; sahanak@gmail.com  
[2] Department of Physics, Virginia Polytechnic Institute and State University, 850 West Campus Drive, Blacksburg, VA 24061, USA  
[3] Carnegie Observatories, Las Campanas Observatory, Colina El Pino, Casilla 601, Chile  
[4] Observatories of the Carnegie Institution for Science, 813 Santa Barbara St., Pasadena, CA 91101, USA  
[5] Institute of Space Sciences (ICE, CSIC), Campus UAB, Carrer de Can Magrans, s/n, E-08193 Barcelona, Spain  
[6] Institut dEstudis Espacials de Catalunya (IEEC), E-08034 Barcelona, Spain  
[7] Homer L. Dodge Department of Physics and Astronomy, 440 West Brooks Street, Room 100, Norman, OK 73019, USA  
[8] Department of Physics and Astronomy, UC Davis, 1 Shields Ave., Davis, CA 95616, USA  
[9] Private Astronomer, USA  
[10] Millennium Institute of Astrophysics, Casilla 36-D, 7591245, Santiago, Chile  
[11] Departamento de Astronomía, Universidad de Chile, Casilla 36-D, Santiago, Chile  
[12] DiRAC Institute, Department of Astronomy, University of Washington, Box 351580, U.W., Seattle, WA 98195, USA  
[13] Department of Physics and Astronomy, Aarhus University, Ny Munkegade, DK-8000 Aarhus C, Denmark  
[14] TMT International Observatory, 100 West Walnut Street, Pasadena, CA 91124, USA  
[15] California Institute of Technology, 1216 E. California Boulevard, Pasadena, CA 91125, USA  
[16] CSIRO Space and Astronomy, PO Box 76, Epping, NSW 1710, Australia  
[17] George P. and Cynthia Woods Mitchell Institute for Fundamental Physics and Astronomy, Department of Physics and Astronomy, Texas A&M University, College Station, TX 77843, USA  
[18] Department of Astronomy, University of Texas, 1 University Station C1400, Austin, TX 78712, USA  
[19] Facultad de Ciencias Astronomicas y Geofisicas, Universidad Nacional de La Plata, Paseo del Bosque S/N, B1900FWA, La Plata, Argentina  

## Abstract

We present multiwavelength time-series spectroscopy of SN 2013aa and SN 2017cbv, two Type Ia supernovae (SNe Ia) on the outskirts of the same host galaxy, NGC 5643. This work utilizes new nebular-phase near-infrared (NIR) spectra obtained by the Carnegie Supernova Project-II, in addition to previously published optical and NIR spectra. Using nebular-phase [Fe II] lines in the optical and NIR, we examine the explosion kinematics and test the efficacy of several common emission-line-fitting techniques. The NIR [Fe II] 1.644 $\mu$m line provides the most robust velocity measurements against variations due to the choice of the fit method and line blending. The resulting effects on velocity measurements due to choosing different fit methods, initial fit parameters, continuum and line profile functions, and fit region boundaries were also investigated. The NIR [Fe II] velocities yield the same radial shift direction as velocities measured using the optical [Fe II] $\lambda$7155 line, but the sizes of the shifts are consistently and substantially lower, pointing to a potential issue in optical studies. The NIR [Fe II] 1.644 $\mu$m emission profile shows a lack of significant asymmetry in both SNe, and the observed low velocities elevate the importance for correcting for any velocity contribution from the host galaxy's rotation. The low [Fe II] velocities measured in the NIR at nebular phases disfavor progenitor scenarios in close double-degenerate systems for both SN 2013aa and SN 2017cbv. The time evolution of the NIR [Fe II] 1.644 $\mu$m line also indicates moderately high progenitor white dwarf central density and potentially high magnetic fields.

*Unified Astronomy Thesaurus concepts:* Type Ia supernovae (1728); Spectroscopy (1558)

## 1. Introduction

Type Ia supernovae (SNe Ia) are powerful cosmological tools. Empirical relations between their bright peak luminosities and light-curve decline rates (e.g., Phillips 1993) allow SNe Ia to be standardizable candles and to be used to map the expansion history of the universe. Observations of distant SNe Ia led to the discovery of the accelerating expansion of the universe (Riess et al. 1998; Perlmutter et al. 1999) and are used to determine the value of $H_0$ (e.g., Freedman et al. 2001).

SNe Ia are widely accepted as the thermonuclear explosions of at least one C–O white dwarf (WD; Hoyle & Fowler 1960). Several explosion mechanisms have been proposed, including scenarios where the explosion is triggered and scenarios where the primary WD approaches the Chandrasekhar mass ($M_{Ch}$) and while the primary WD is substantially sub-Chandrasekhar in mass (sub-$M_{Ch}$). Separately, possible progenitor systems may be categorized as single or double degenerate, with the companion star being a nondegenerate star or a WD, respectively.

For example, the merging of two WDs caused by the orbital decay of a double-degenerate system with a combined mass near $M_{Ch}$ (Webbink 1984) can lead to an SN Ia explosion resulting in the destruction of both WDs (Iben & Tutukov, 1984). Moreover, head-on collisions of WDs have been proposed as a way to produce a shock-triggered thermonuclear explosion (Rosswog et al. 2009). This scenario predicts doubly peaked nebular-phase emission lines owing to an underlying bimodal velocity distribution from the kinematics of the two WDs (Dong et al. 2015).

In the $M_{Ch}$ scenario, the nuclear flame front undergoes a deflagration-to-detonation transition (DDT). The initial subsonic deflagration allows pre-expansion and the production of intermediate-mass elements prevalent in SN Ia spectra, before the ensuing detonation. DDT models have generally been successful at reproducing a wide range of observed properties. It is currently unclear how and whether the transition from deflagration to detonation can occur in nature, but turbulence may play a role (Poludnenko et al. 2019). Two-dimensional (e.g., Niemeyer & Hillebrandt 1995; Lisewski et al. 2000) and three-dimensional calculations (e.g., Livne & Arnett 1993; Plewa et al. 2004) for deflagration fronts predict strong mixing by Rayleigh–Taylor instabilities, but these results appear to be at odds with the observed layered structure. However, the presence of magnetic fields may suppress large-scale instabilities within the deflagration front (Hristov et al. 2018).

The helium-detonation or double-detonation scenario consists of a sub-$M_{Ch}$ primary WD with a thin surface helium shell accreted from a companion star. The surface helium shell detonates and drives a shock wave inward, igniting carbon in the interior of the WD (Woosley & Weaver 1994). This scenario, which can explode WDs with a range of masses, may provide the most natural explanation for the observed range of peak luminosities in SNe Ia (e.g., Shen et al. 2021a, 2021b).

Although the $M_{Ch}$ and sub-$M_{Ch}$ explosion scenarios may occur in either single- or double-degenerate progenitor systems, specific combinations of progenitor systems and explosion mechanisms have also been explored. One such example is the "dynamically driven double-degenerate double-detonation" ($D^6$) scenario (Shen et al. 2018a). In this scenario, the surviving companion WD is expelled out of its binary orbit and becomes a hypervelocity runaway WD, moving at speeds comparable to their pre-explosion orbital velocities, which are typically at least 1000 km s$^{-1}$ (Shen et al. 2018b). Comparably high SN remnant velocities would also be expected in the nebular-phase spectra.

Nebular-phase spectra can reveal the kinematics, geometry, and chemical composition of the innermost regions of the SN, and the time evolution of specific emission features can be used to study the underlying physics. For example, the width of nebular-phase optical Fe lines correlates with light-curve decline rate (Mazzali et al. 2007), and the evolution and velocity shifts of several prominent emission features have been used to study asymmetries and viewing angle effects (Maeda et al. 2010a). Furthermore, SN Ia ejecta become optically thin at NIR wavelengths much earlier than at optical wavelengths, so optical and NIR spectra taken around the same time can be used to examine different regions of the ejecta (Wheeler et al. 1998).

Currently, there are significantly more optical nebular-phase spectra in the literature than NIR spectra. As such, much of our current understanding of the nebular phase of SNe Ia is based on optical features. The deflagration products and their velocities can be measured at nebular phases using optical [Fe II] and [Ni II] lines owing to emission from the innermost region of the SN (Maeda et al. 2010b). Similarly, the velocities of detonation products can be measured through optical [Fe III] features, and in particular the [Fe III] $\lambda$4701 line. The strong emission feature at $\sim$7200 Å is a blend of iron, nickel, and possibly calcium lines. It is difficult to isolate the individual components, but it is assumed that the [Fe II] $\lambda$7155 and [Ni II] $\lambda$7378 lines dominate this feature. This blended feature is commonly used to assess the line-of-sight velocity of the deflagration ash (Maeda et al. 2010a). The average of the velocities of these two lines gives $v_{neb}$, which has been shown to correlate with the high velocity gradient (HVG) and low velocity gradient (LVG) classifications (Maeda et al. 2010a; Graham et al. 2017).

In the NIR, the [Fe II] 1.644 $\mu$m emission line is well isolated and remains strong throughout the nebular phase. Unlike optical [Fe II] lines, the lack of line blending makes this NIR [Fe II] line an excellent gauge of the conditions within the central region of the SN. Diamond et al. (2018) use the width of this emission line in SN 2014J to conclude a relatively low progenitor WD central density in the context of a $M_{Ch}$ DDT scenario. The [Fe II] 1.644 $\mu$m line can also be used to study geometric effects within the SN ejecta. Diamond et al. (2018) used this to examine the distribution of material in the inner layers of the SN and test the likelihood of magnetic fields with a range of possible strengths.

Previous studies, such as Maguire et al. (2018), use the blended [Fe II] feature near $\sim$1.2 $\mu$m, as opposed to the [Fe II] 1.644 $\mu$m line, and opt for a multi-Gaussian fit to obtain nebular-phase NIR [Fe II] velocities. The $\sim$1.2 $\mu$m region is dominated by multiple [Fe II] lines (Flörs et al. 2020) and should provide similar kinematic information to the [Fe II] 1.644 $\mu$m line because the emission is expected to have the same origin (Maguire et al. 2018).

In this work, we present NIR and optical time-series spectroscopy of the "siblings" SN 2013aa and SN 2017cbv at nebular phases. These two SNe Ia exploded in the same host galaxy, NGC 5643, roughly 4 yr apart (Burns et al. 2020). They provide an excellent opportunity to investigate intrinsic properties of these SNe Ia, particularly those that are largely independent of distance and reddening effects. The observations are outlined in Section 2. The method for quantifying the spectroscopic features is presented in Section 3. The results and their implications are discussed in Section 4. Finally, the conclusions are summarized in Section 5.

## 2. Observations

Both SN 2013aa and SN 2017cbv were well observed at early times by the Carnegie Supernova Project-II (CSP-II; Hsiao et al. 2019; Phillips et al. 2019) and others. They exploded at the outskirts of NGC 5643, a nearby galaxy with its distance determined by primary methods, such as the tip of the red giant branch. These two SNe Ia show remarkable similarities in both their spectroscopic and photometric properties at early phases (Burns et al. 2020), and several studies have examined these SNe Ia individually at nebular phases (e.g., Graham et al. 2017; Sand et al. 2018). Nebular-phase spectra are often used to search for evidence of a nondegenerate companion star, as residual material such as H may be embedded in the low-velocity ejecta (Marietta et al. 2000). Neither SN 2013aa (Graham et al. 2017) nor SN 2017cbv (Wang et al. 2020; Sand et al. 2021) exhibits evidence of surviving material from a nondegenerate companion.

### 2.1. Early Phase Properties

SN 2013aa was initially classified as 91T-like (Jacobson-Galán et al. 2018) but was subsequently shown to be a normal SN Ia at the bright and slowly declining end of the normal population. It was followed by the CSP-II as part of the "Physics subsample" (Phillips et al. 2019) with nightly photometric observations at early phases, as well as time-series NIR spectroscopy from peak brightness to nebular phases (Hsiao et al. 2019). SN 2013aa reached peak brightness on MJD $56{,}343.20 \pm 0.07$ days in the $B$ band. All light-curve parameters were adopted from Burns et al. (2020) and directly measured via spline fits since the early light curves are densely sampled.

SN 2017cbv was discovered by the Distance Less Than 40 Mpc survey (DLT40; Valenti et al. 2017; Tartaglia et al. 2018) exceptionally close to the time of explosion. The early light curves revealed a photometric "blue bump" that may be a signature of impact with a nondegenerate companion (Hosseinzadeh et al. 2017). It is unknown whether SN 2013aa also exhibited a similar blue excess, since the light-curve coverage does not extend to such early times. SN 2017cbv reached peak brightness on MJD $57{,}840.54 \pm 0.15$ days. It also showed strong C II $\lambda 6580$ absorption in the early spectra, indicating the presence of unburned material from the progenitor system (Hosseinzadeh et al. 2017).

SN 2013aa and SN 2017cbv have nearly identical light-curve decline rates of $\Delta m_{15}(B) = 0.95 \pm 0.01$ mag and $0.96 \pm 0.02$ mag, respectively. These are slightly slower decline rates when compared to the average of the normal SNe Ia in the larger sample presented in Phillips et al. (2019). SN 2013aa and SN 2017cbv also have nearly identical peak $B$-band magnitudes: $11.094 \pm 0.003$ mag and $11.118 \pm 0.011$ mag, respectively. Both SNe also exhibit a subtle $i$-band "kink" (Pessi et al. 2022), which further demonstrates the close resemblance of these sibling SNe at early times.

Furthermore, they show remarkable similarities in their spectral properties. From their optical spectra, they are situated at the same location in the Branch diagram (Branch et al. 2006; Burns et al. 2020). Their early-time NIR spectra imply virtually the same $^{56}$Ni production and distribution (Burns et al. 2020) using the $H$-band break (Hsiao et al. 2013; Ashall et al. 2019).

### 2.2. Nebular-phase Spectroscopy

One of the main goals of this work is to compare the kinematics of the remnants of two SNe Ia obtained using both optical and NIR spectral features. Specifically, we emphasize the use of the well-isolated NIR [Fe II] 1.644 $\mu$m emission line and the techniques shown in Diamond et al. (2015). Both SN 2013aa and SN 2017cbv have other previously published nebular-phase spectra that we include to allow for multi-wavelength time-series analysis.

The previously unpublished data come from CSP-II, a 4 yr NSF-funded program that obtained follow-up observations of SNe Ia in both the optical and the NIR (Phillips et al. 2019), using resources at the Las Campanas Observatory (LCO). Six nebular-phase NIR spectra were obtained with the Folded-port InfraRed Echelette (FIRE; Simcoe et al. 2013) on the 6.5 m Magellan Baade telescope at LCO. These spectra were obtained using the conventional ABBA nod-along-the-slit technique in the high-throughput prism mode with a $0\rlap.{''}6$ slit and a wavelength coverage of 0.8–2.5 $\mu$m. They were reduced and telluric corrected following the methods described by Hsiao et al. (2019). We did not adopt Poisson errors for the flux uncertainties; rather, they were measured at each pixel using the dispersion in the counts of the multiple ABBA exposures. Each spectrum has at least 16 ABBA exposures. The three FIRE spectra obtained for each of SN 2013aa and SN 2017cbv compose a nebular-phase data set with an approximate cadence of 2–3 months from $\sim +300$ to $\sim +500$ days past peak brightness.

When possible, we coordinated complementary optical observations at approximately the same phases as the FIRE spectra. These optical spectra were obtained with the Gemini Multi-Object Spectrograph (GMOS; Hook et al. 2004) on the 8.1 m Gemini South telescope and were reduced using the method presented in Graham et al. (2017). Two GMOS spectra of SN 2013aa are presented in Graham et al. (2017), and one GMOS spectrum of SN 2017cbv is presented in Graham et al. (2022). Finally, we include two XShooter spectra of SN 2013aa published by Maguire et al. (2018) and one LDSS-3 spectrum of SN 2017cbv published by Tucker et al. (2020). The full data set used in our analysis is presented in Table 1 and Figure 1.

### 2.3. Host Galaxy Recession Velocities

Since the host galaxy NGC 5643 is nearby and radial velocity maps are available, we opted to correct the observed spectra to the rest frames of SN 2013aa and SN 2017cbv using the recession velocities at the sites of the explosions rather than using a single systemic velocity for the entire galaxy.

Integral field spectroscopy of NGC 5643 (Erroz-Ferrer et al. 2019) was obtained with the Multi-Unit Spectroscopic Explorer (MUSE; Bacon et al. 2010), mounted to the Unit 4 telescope (UT4) at the Very Large Telescope (VLT). An H$\alpha$ velocity map of the central region of NGC 5643 was then constructed to reveal that NGC 5643 is not completely face-on (Galbany et al. 2016). The NW and SE halves of the spiral galaxy are redshifted and blueshifted, respectively, relative to the systemic recession velocity. However, the observed field was not wide enough to cover the sites of SN 2013aa and SN 2017cbv.

Radio observations of NGC 5643 using the H I 21 cm spectral line were obtained with the Australia Telescope Compact Array (ATCA). For a description of the data reduction and imaging see Koribalski et al. (2018). The H I velocity map covers the locations of both SN 2013aa and SN 2017cbv and is shown in Figure 2. The velocity resolution is 4 km s$^{-1}$, and the beam size is approximately $80''$. Using the velocity map, we estimated the recession velocities at the sites of SN 2013aa and SN 2017cbv to be $1166 \pm 19$ km s$^{-1}$ and $1264 \pm 21$ km s$^{-1}$, respectively. The estimated uncertainties are derived from the beam size of the radio observations, as well as stellar velocity dispersion measured using MUSE data. The recession velocities at the sites of SN 2013aa and SN 2017cbv are slightly blueshifted and redshifted, respectively, relative to the systemic velocity of 1199 km s$^{-1}$ listed in the NASA/IPAC Extragalactic Database (NED) (2019) from Koribalski et al. (2004).

### 3. Spectroscopic Measurements

In this section, we present the methods for quantifying the properties of nebular-phase forbidden emission lines and their associated uncertainties. In particular, the optical [Fe II] $\lambda 7155$ and the NIR [Fe II] 1.644 $\mu$m lines were studied and compared

**Table 1**
Nebular-phase Optical and NIR Spectra of SN 2013aa and SN 2017cbv

| SN | Phase (days)[a] | MJD | Telescope + Instrument | Optical or NIR | Source |
| --- | --- | --- | --- | --- | --- |
| SN 2013aa | +362 | 56,696.4/56714.4[b] | VLT + XShooter | Optical + NIR | Maguire et al. (2018) |
| SN 2013aa | +368 | 56,710.8 | Magellan Baade + FIRE | NIR | This work |
| SN 2013aa | +400 | 56,743.4 | Gemini South + GMOS | Optical | Graham et al. (2017) |
| SN 2013aa | +426 | 56,769.2 | VLT + XShooter | Optical + NIR | Maguire et al. (2018) |
| SN 2013aa | +428 | 56,770.8 | Magellan Baade + FIRE | NIR | This work |
| SN 2013aa | +497 | 56,840.0 | Gemini South + GMOS | Optical | Graham et al. (2017) |
| SN 2013aa | +505 | 56,848.6 | Magellan Baade + FIRE | NIR | This work |
| SN 2017cbv | +309 | 58,149.7 | Magellan Baade + FIRE | NIR | This work |
| SN 2017cbv | +317 | 58,158.0[c] | Magellan-Clay + LDSS-3 | Optical | Tucker et al. (2020) |
| SN 2017cbv | +434 | 58,274.7 | Magellan Baade + FIRE | NIR | This work |
| SN 2017cbv | +466 | 58,307.0[c] | Gemini South + GMOS | Optical | Graham et al. (2022) |
| SN 2017cbv | +510 | 58,350.5 | Magellan Baade + FIRE | NIR | This work |

**Note.**
[a] Phase relative to $B$-band maximum from Burns et al. (2020).
[b] This XShooter spectrum is a combination of observations obtained on two different nights and was first published in Maguire et al. (2016), where the phase was reported as the midpoint between the two observation dates. We use this same midpoint to determine the phase relative to the $B$-band maximum from Burns et al. (2020).
[c] Observation date was published as UT date, which we then converted to MJD.

in detail. In order to assess the robustness of each measurement, the effects of varying the fit region boundary, continuum, and profile function selections were also examined.

### 3.1. Optical [Fe II] λ7155 Velocity

The [Fe II] λ7155 emission line forms one of the strongest [Fe II] features in the optical and is most often used to measure the velocity shifts of SN Ia remnants. Despite its prevalent use, this line is heavily blended with neighboring [Fe II], [Ni II], and [Ca II] lines (e.g., Mazzali et al. 2015).

The region around the optical [Fe II] λ7155 emission line (approximately 7000–7600 Å) most often manifests as a double-peaked feature. Thus, Maeda et al. (2010a) adopted the simple two-Gaussian fit to determine the velocity shift, attributing the blue-side emission to [Fe II] λ7155 and the red-side emission to [Ni II] λ7378. More recently, a six-Gaussian fit method emerged in an attempt to account for the line blending (e.g., Maguire et al. 2018; Graham et al. 2022). The six components included are four [Fe II] lines (λλ7155, 7172, 7388, and 7453) and two [Ni II] lines (λλ7378 and 7412).

Here, a similar method to that of Maguire et al. (2018) was adopted: fixing the relative strengths and velocities for lines of the same ion. The relative strengths of the [Fe II] and [Ni II] lines were fixed using the same atomic data as in Höflich (2009) and Diamond et al. (2015) and references therein. In this section, both the two-Gaussian and six-Gaussian fitting methods are presented. The best-fit parameters were determined via nonlinear least-squares fitting using the package LMFIT (Newville et al. 2014).

In most previously published works, a continuum determined by a straight line connecting the manually defined boundaries of the emission feature is first removed before the fitting process (e.g., Graham et al. 2017; Maguire et al. 2018). The boundaries are manually chosen to mitigate the effects of neighboring emissions in heavily blended regions. In this work, we opted to include a continuum in the overall profile fit rather than removing one beforehand. Furthermore, allowing a nonzero slope for a linear continuum, as opposed to a fixed flat continuum, was also shown to yield better fits without overfitting, as assessed by the reduced $\chi^2$. Thus, we also adopted a linear continuum in the fitting process.

In the optical, choosing a nonzero-slope continuum as opposed to a flat continuum causes an average difference in the resulting velocity shifts of 189 km s$^{-1}$. This result is independent of the choice of the profile function. Furthermore, the Gaussian fits with linear continuum consistently resulted in redshifted velocities compared to the Gaussian fits with a flat continuum.

To estimate the velocity uncertainty during the fitting process, three sources were considered: the choice of the fit boundaries, the choice of the initial profile center, and the flux error of the observed spectrum. The effects of these uncertainty sources were simulated by generating 1000 realizations of the observed spectra with random boundaries (uniform distribution) within 20 Å of the original choice, random initial line centers (uniform distribution) within 50 Å of the best-fit value, and random flux fluctuations produced by a normal distribution with the width set to the measured flux error. A velocity shift is then obtained for each realization, and the standard deviation of all measured velocities was adopted as the velocity uncertainty.

The two-Gaussian and six-Gaussian best-fit profiles are presented in Figures 3 and 4, respectively. The two methods show similar goodness of fit (corroborated by similar reduced $\chi^2$ values for the two-Gaussian and six-Gaussian fits), indicating that increasing the number parameters does not substantially improve the fits. Furthermore, six-Gaussian fits are susceptible to the choice of the initial parameters, as the large number of parameters can cause the $\chi^2$ minimizer to find the local minima rather than the best solution.

The +497-day optical spectrum of SN 2013aa has poor fit results for both optical methods owing to an emission feature emerging between the double peaks (most evident in Figure 4). The feature is most likely formed by [Ca II] λλ7291 and 7324. Furthermore, most optical spectra in our sample show an emerging emission feature located on the red side of this double-peaked region of interest. This emerging feature may be attributed to the [Fe II] λ7638 line (Mazzali 2015) and can affect the red wing of the multi-Gaussian fit.

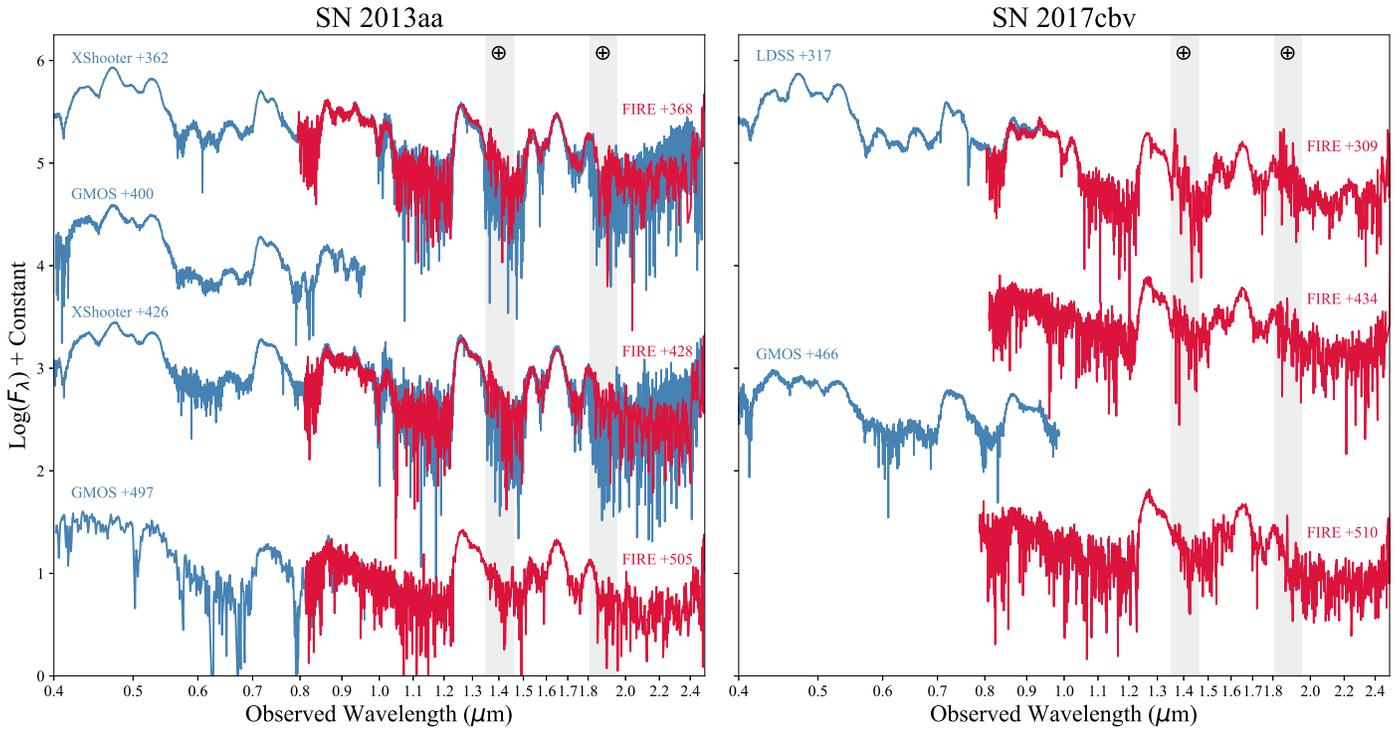

**Figure 1.** Optical and NIR nebular-phase spectra of SN 2013aa and SN 2017cbv. Spectra taken at similar phases are shown together, providing multiwavelength snapshots of these two SNe at late times. The instrument and phase relative to *B*-band maximum are noted for each spectrum. The gray vertical bands indicate the regions of heavy telluric absorption in the NIR. The spectra are presented in observed wavelength shown in log scale. Negative fluxes are excluded for presentation but are included in the analyses, as they are due to noise in the observed spectra.

(The data used to create this figure are available.)

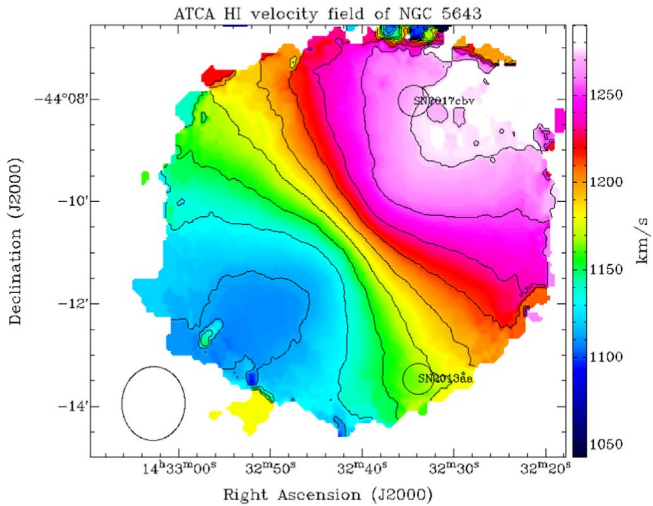

**Figure 2.** Radio observations from ATCA showing the H I velocity field of NGC 5643 (Koribalski et al. 2018). The locations of SN 2013aa and SN 2017cbv are labeled and show that these sibling SNe require different velocity corrections to obtain more precise rest-frame velocity measurements. The beam size is also indicated in the lower left corner.

Different profile functions were also tested. Changing from a Gaussian to a Voigt function yielded differences within $1\sigma$ for most spectra in the sample. In general, the choice of line profile and continuum does not have as significant of an effect on the measured line velocities as other factors such as flux error and choice of initial parameter values.

The resulting velocity measurements are presented in Table 2. The final velocity uncertainty estimates include the following components: the robustness of the fitting process as described above, the least-squares fit parameter uncertainty, and the uncertainty in the recession velocity (Section 2.3). The two-Gaussian and six-Gaussian fits yield consistent results for both SNe. The differences in the velocities measured by the two different methods are generally within $1\sigma$. On average, the measured velocity shifts have a difference of 398 km s$^{-1}$ for SN 2013aa and 375 km s$^{-1}$ for SN 2017cbv. However, the two-Gaussian fits yield higher velocity measurements in five out of the six optical spectra in our sample, pointing to a potential systematic effect.

### 3.2. NIR [Fe II] 1.644 μm Velocity

The NIR [Fe II] 1.644 μm line is a well-isolated feature with minimal blending from neighboring emission lines (e.g., Höflich et al. 2004; Diamond et al. 2015). Visual inspections of this feature in our sample showed no evidence of significant contamination of neighboring [Fe II] and [Fe III] features (Figure 5 of Diamond et al. 2018), asymmetry (e.g., Hoeflich et al. 2021), or multiple components in the context of a direct collision scenario (e.g., Dong et al. 2015; Mazzali et al. 2018). A possible exception is the last NIR spectrum of SN 2017cbv, which shows a slight asymmetry only near the peak of the feature, but the spectrum also has more noise in this region than the other spectra in our NIR sample. Thus, we elected to fit the strong emission feature with a single-profile function via nonlinear least-squares fitting, assuming that the [Fe II] 1.644 μm line is dominant.

As in the optical, we only consider the case where a continuum is included in the least-squares fit and not the case where a continuum is first removed before the fit. Two options for the continuum were tested: a linear continuum allowing for a nonzero slope and a flat continuum fixing the slope at zero. For comparison, past studies of the NIR [Fe II] 1.644 μm line assumed either a flat continuum with contribution guided by models (Diamond et al. 2018) or no continuum (Dhawan et al. 2018).

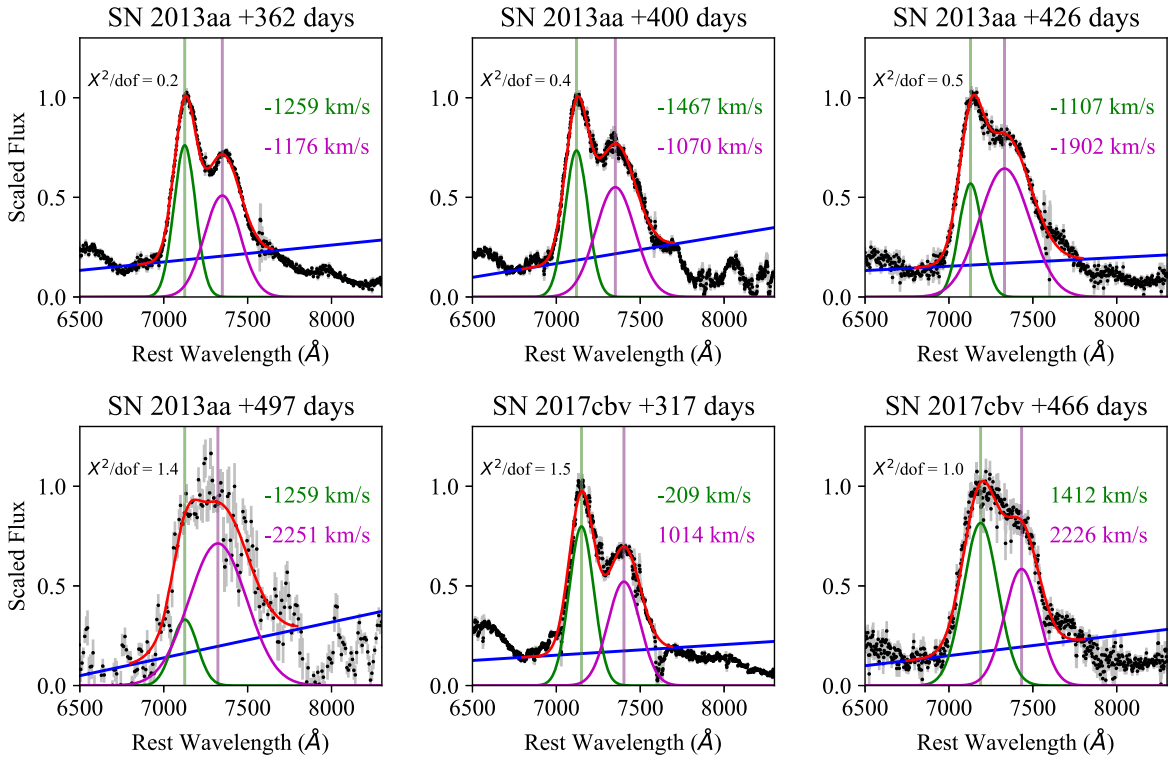

**Figure 3.** Optical two-Gaussian fits used to measure line velocity shifts of the [Fe II] λ7155 feature (shown in green) and the [Ni II] λ7378 feature (shown in magenta). The two Gaussians are simultaneously fit with a linear continuum (blue), with the best-fit function shown in red. The +497-day spectrum of SN 2013aa is not well fit by the two-Gaussian function, likely due to an emerging [Ca II] feature. The reduced $\chi^2$ per degree of freedom shows that this two-Gaussian method can easily underfit the data and indicates that two Gaussians may not be enough to construct this blended feature.

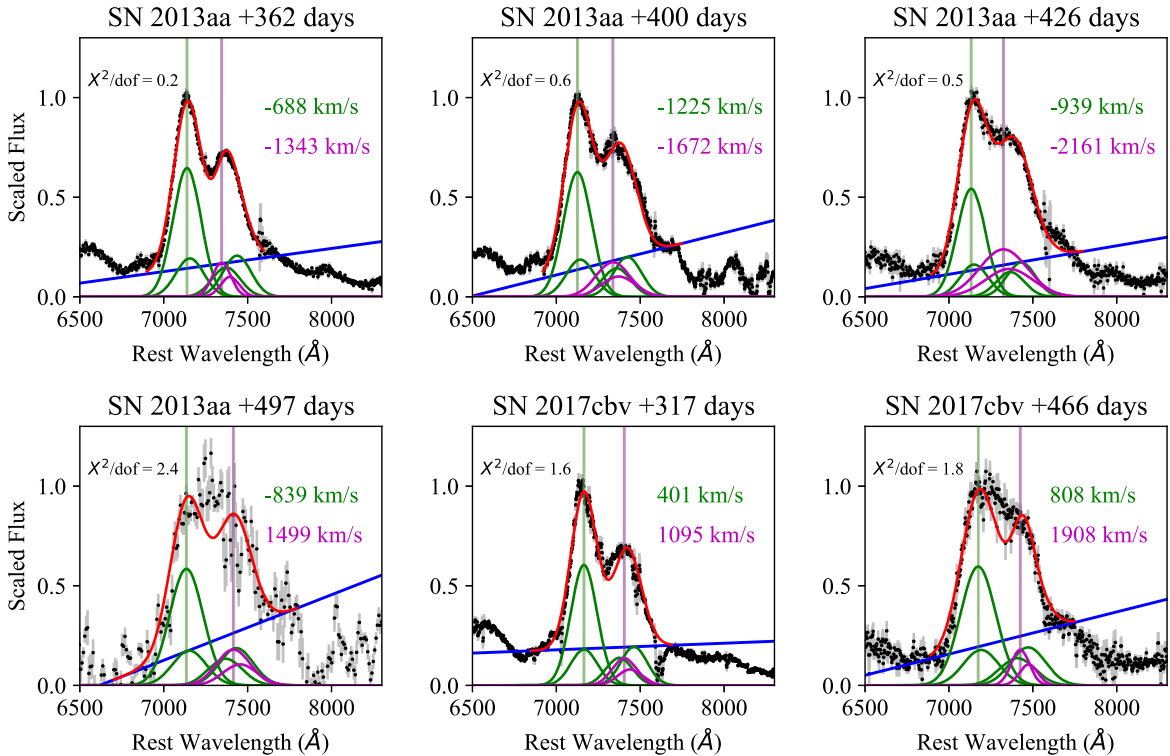

**Figure 4.** Profile fits to the wavelength region near the optical [Fe II] λ7155 line. Each fit shown is a combination of six Gaussian functions, representing four [Fe II] lines (green) and two [Ni II] lines (magenta), combined with a linear continuum. The best-fit continuum and final profile are shown in blue and red, respectively. The resulting line-center velocity shift is labeled for each spectrum.

Table 2
Rest-frame Optical [Fe II] λ7155 and [Ni II] λ7378 Line Velocities Obtained Using Least-squares Fits of Multiple Gaussian Profiles and a Linear Continuum

| SN | Phase (days) | Two-Gaussian Fit [Fe II] Velocity (km s$^{-1}$) | [Ni II] Velocity (km s$^{-1}$) | Six-Gaussian Fit [Fe II] Velocity (km s$^{-1}$) | [Ni II] Velocity (km s$^{-1}$) |
|---|---|---|---|---|---|
| SN 2013aa | +362 | −1259 ± 741 | −1176 ± 435 | −688 ± 299 | −1343 ± 1127 |
| SN 2013aa | +400 | −1467 ± 567 | −1070 ± 430 | −1225 ± 623 | −1672 ± 799 |
| SN 2013aa | +426 | −1107 ± 758 | −1902 ± 1015 | −939 ± 615 | −2161 ± 547 |
| SN 2013aa | +497 | −1259 ± 747 | −2251 ± 974 | −839 ± 687 | 1499 ± 2616 |
| SN 2017cbv | +317 | −209 ± 411 | 1014 ± 567 | 401 ± 709 | 1095 ± 1159 |
| SN 2017cbv | +466 | 1412 ± 940 | 2226 ± 613 | 808 ± 770 | 1908 ± 1329 |

Next, two functions were considered for the single-profile fit: Gaussian and Voigt functions. Overall, the determination of the line centers was stable regardless of the choice of the continuum and profile function, and there was no evidence of a systematic shift in the line centers from one method to the next.

The reduced $\chi^2$ of the region near the center are slightly over 1 for most spectra, indicating that the choice of a one-profile fit is adequate for measuring the line center. Furthermore, the reduced $\chi^2$ does not vary drastically between different combinations of continuum and profile function. We chose to present the fit results from a single-Gaussian function in combination with linear continuum fits as the final measurements. These fits are shown in Figure 5 and the average reduced $\chi^2$ values are listed for each fit method in Table 4.

The velocity uncertainty was estimated in the same fashion as the optical fit (detailed in Section 3.1). The effects of fit boundary choice and choice of the initial parameters are again simulated. As expected, the determination of the best-fit parameters for the NIR [Fe II] 1.644 μm is robust against varying choices of initial parameters. This is in contrast to the six-Gaussian fit for the region near the optical [Fe II] λ7155 line, where there are often degeneracies in the best-fit parameters. The final uncertainty estimates again include the robustness of the fitting process, the least-squares fit parameter uncertainty, and the uncertainty in the host recession velocity. The velocity measurements and associated uncertainties are presented in Table 3.

Unlike the optical, testing different continuum options in the NIR yields different results that are affected by the choice of line profile function. For a Gaussian line profile, the linear (nonzero-slope) continuum and flat (zero-slope) continuum combinations resulted in an average difference of 107 km s$^{-1}$ for SN 2013aa and 136 km s$^{-1}$ for SN 2017cbv. A Voigt line profile was less affected by the change in continuum, resulting in an average velocity difference of 18 km s$^{-1}$ for SN 2013aa and 44 km s$^{-1}$ for SN 2017cbv. This may be attributed to how well the combination of the continuum and line profile fits the wings of the emission feature. Overall, the difference in line profile had the most significant effect in the NIR when paired with a linear (nonzero-slope) continuum. On average, these differences are 130 and 184 km s$^{-1}$ for SN 2013aa and SN 2017cbv, respectively. This may again be attributed to how well the combination of the continuum and line profile fits the wings of the emission feature.

For all continuum types, the NIR fits generally agreed with model predictions of the strength of the continuum with respect to the peak of the [Fe II] 1.644 μm line (Diamond et al. 2015). The models estimate a continuum level of ∼10% at +300 days, 7.5% at +400 days, and 5% at +500 days.

### 3.3. NIR [Fe II] 1.644 μm Line Width

The width of the NIR [Fe II] 1.644 μm line has been used to estimate the initial WD central density and magnetic field (Diamond et al. 2015, 2018). The line width was measured following the same procedure as in Diamond et al. (2018). Briefly, the line width was measured at a chosen flux scale relative to the maximum, in this case 0.6 of the peak. The relative flux height of 0.6 was chosen because it maximized the separation between the different DDT models. Following the method of Diamond et al. (2018), a flat continuum was removed before the measurement. As the choice of the continuum directly affects the width measurement, we considered two extremes. In the extreme of no continuum subtraction, the width measured represents the maximum value. At the other extreme, the maximum continuum level is determined by the flux minimum of the red side of the [Fe II] 1.644 μm feature. Since this extreme likely represents an oversubtraction of continuum, the width measured yields the minimum value. The two extremes give the possible range of the line width. The results are presented and discussed in Section 4.5.

## 4. Discussion

The time-series nebular-phase optical and NIR spectroscopy of two SNe Ia exploding in the same host galaxy offers an opportunity to study their explosion kinematics and symmetry, as well as their progenitor central densities and possible magnetic fields. The main results are discussed in the following subsections.

### 4.1. Fit Method Results

Complementary optical and NIR data allowed for the comparison of the velocity measurements from two methods: (1) fitting a single or multiple [Fe II] lines in the optical, including the widely used [Fe II] λ7155 line and (2) fitting a single NIR [Fe II] 1.644 μm line. As the phases observed (past +300 days) are generally considered nebular or close to nebular, the expectation is that the optical and NIR lines of the same ion should yield the same velocity.

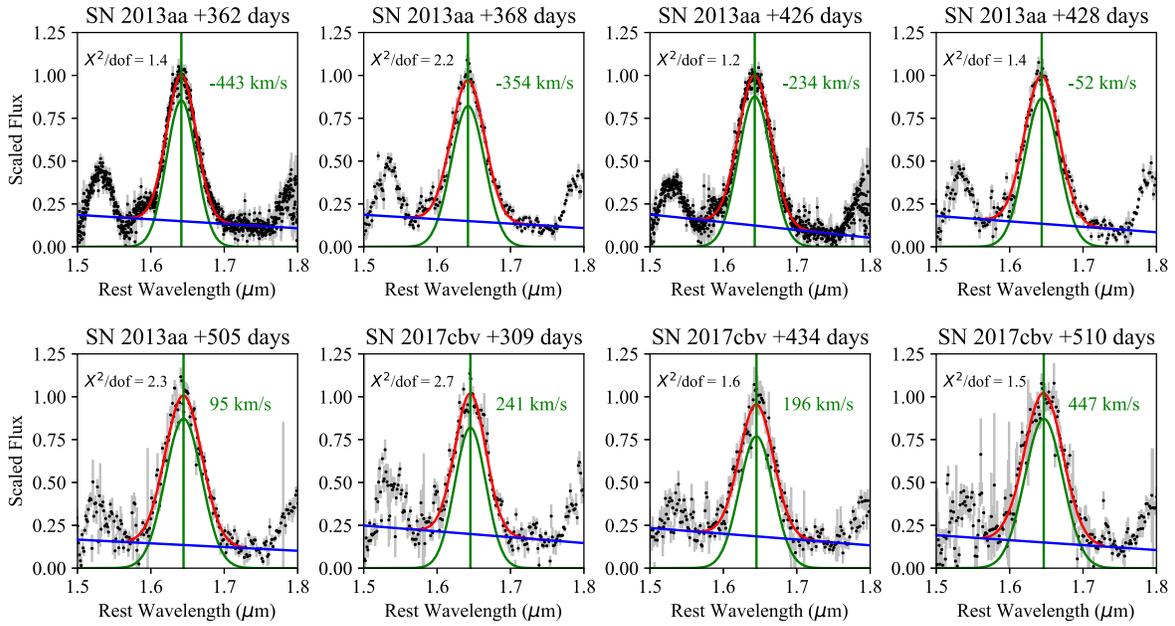

**Figure 5.** Results of fitting each observed NIR [Fe II] 1.644 μm spectral feature with a combination of a single-Gaussian profile and a linear continuum, which allows for a nonzero slope. The Gaussian function, continuum, and final best-fit profile are in green, blue, and red, respectively. The resulting [Fe II] line velocity is also labeled for each spectrum.

Table 3
Measured [Fe II] 1.644 μm Velocity

| SN | Phase (days) | Velocity (km s$^{-1}$) |
|---|---|---|
| SN 2013aa | +362 | −443 ± 67 |
| SN 2013aa | +367 | −354 ± 89 |
| SN 2013aa | +426 | −234 ± 103 |
| SN 2013aa | +427 | −52 ± 127 |
| SN 2013aa | +505 | 95 ± 130 |
| SN 2017cbv | +309 | 241 ± 175 |
| SN 2017cbv | +434 | 196 ± 147 |
| SN 2017cbv | +510 | 447 ± 178 |

**Note.** Rest-frame [Fe II] 1.644 μm line velocities from the least-squares fit of a single-Gaussian profile and a linear continuum allowing for a nonzero slope. The [Fe II] velocities are consistently lower in the NIR compared to the [Fe II] velocities measured by both optical methods. This discrepancy is further investigated in Section 4.1.

Table 4
Comparison of Fit Methods

| Fit Method | Average $\chi^2$/dof |
|---|---|
| Optical two-Gaussian + nonzero-slope continuum | 0.83 |
| Optical six-Gaussian + nonzero-slope continuum | 1.18 |
| NIR Gauss + zero-slope continuum | 1.78 |
| NIR Gauss + nonzero-slope continuum | 1.78 |
| NIR Voigt + zero-slope continuum | 1.70 |
| NIR Voigt + nonzero-slope continuum | 1.76 |

**Note.** The average $\chi^2$ per degree of freedom are computed using the best fit results of methods using different combinations of line profile functions and continuums. Based on these $\chi^2$/dof values, the six-Gaussian fit method provides a better overall fit to the blended feature that includes the optical [Fe II] λ7155 \AA line. The optical two-Gaussian fit method consistently overfits the data when compared to the other methods that were examined. In the NIR, the average $\chi^2$/dof values are similar for all combinations of line profile function and continuum type. This indicates that the choice of fit method does not affect the resulting line velocity measurements in the NIR as significantly as in the optical.

The Doppler shift velocities obtained in the optical and the NIR are compared in Figure 7. The first thing of note is that the NIR measurements are systematically lower in the absolute sense than the optical ones. In the optical, the six-Gaussian measurements are systematically lower than those measured using two-Gaussian profiles. While the optical measurements generally provide the same shift directions as the NIR, the absolute velocity values are systematically higher by several hundreds of kilometers per second compared to the NIR (~600 km s$^{-1}$ on average). If the uncertainty estimates are correct, the NIR measurements appear to be accurate enough to show the [Fe II] velocity evolving to close to 0 km s$^{-1}$ at ~+500 days for SN 2013aa.

To estimate the velocity uncertainties, we considered multiple sources of error that may have been neglected in previous works. These included the robustness of the fitting process, the least-squares fit parameter uncertainty, and the uncertainty in the host recession velocity. For the robustness of the fitting process, the effects of the fit boundary choice and choice of the initial parameters were simulated in the Monte Carlo scheme. The uncertainty associated with the fitting process is by far the dominant source at one order of magnitude larger in velocity than the other two mentioned sources combined.

The velocity uncertainties in the optical measurements are consistently higher than the NIR ones (on average 6 times higher). Thus, the NIR measurements are, by comparison, much more robust against varying choices of fit boundaries and initial parameters. This result is perhaps not surprising given the heavy line blending and the high number of fit parameters in the optical, even when the width, line ratio, and velocity corresponding to the same ion are fixed.

We also tested several different possible continuum types, such as a flat continuum and a linear continuum that allows for a nonzero slope. To maintain consistency in our comparison methods, we include the continuum in the fitting process as opposed to subtracting a continuum before performing the fit. The latter approach was employed by Maguire et al. (2018) and Graham et al. (2017), who first subtracted a linear continuum joining two manually chosen end points. Manually choosing the red-side boundary can mitigate the effects of neighboring lines. One other difference between this work and previously published studies is the condition that the continuum line is not required to pass through the two end points of the fit region. It is a subtle difference but is necessary to consider because the results show that the boundary of the fit region can affect the measured line velocities. The choices of line profile and continuum can also influence the measured line velocities, but these effects are small compared to those detailed in Section 3.

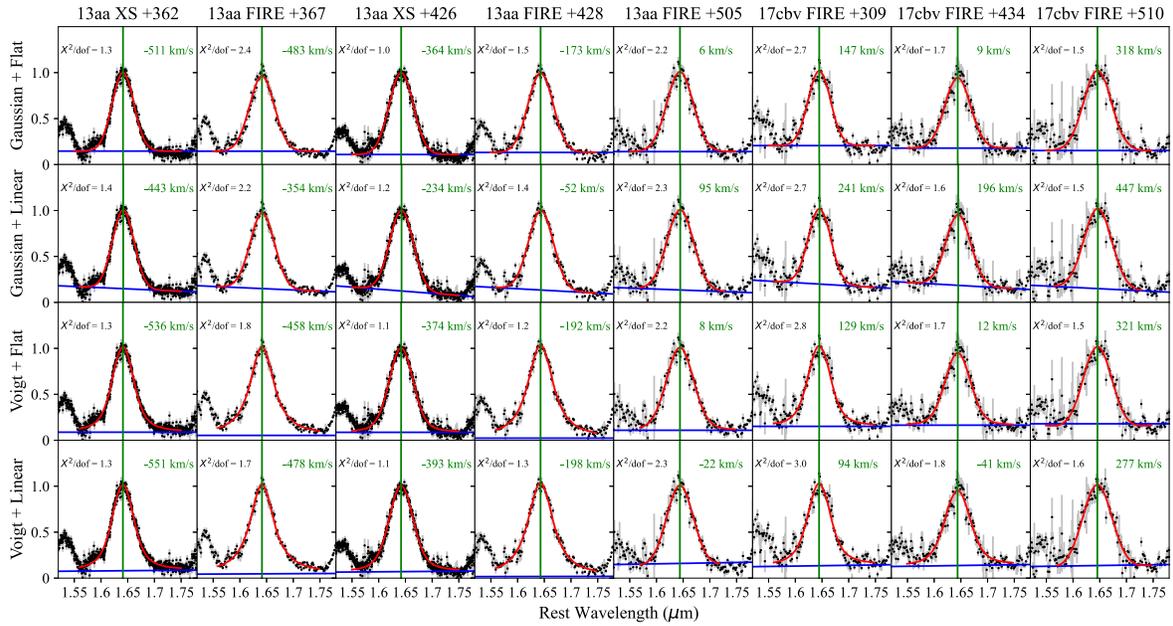

**Figure 6.** Comparison of NIR fit methods using different line profile function and continuum combinations. Similar comparisons were conducted for the optical multi-Gaussian fits, and these combinations of line profiles and continuum types are used to assess the robustness of each fit method. We conclude that the NIR [Fe II] 1.644 $\mu$m feature provides the most robust line velocity measurements, as the choice of fit method does not affect the resulting velocity measurements as drastically as in the optical.

Another factor that could affect the outcome of the velocity measurement is changing or emerging features in the fit region. In Section 3, the main contaminants that could affect the profile shapes in both the optical and the NIR were identified.

In the optical, the feature composed of [Ca II] $\lambda\lambda$7291 and 7324 is present at the center of the double-peaked Fe/Ni feature, between the two peaks. In the last optical spectrum of SN 2013aa at +497 days, the [Ca II] feature is clearly present, but these lines are not included in the fits of most studies. Since the Ca feature is at the center of the profile fit, it would have a significant effect on the determination of the profile center.

Even though the [Fe II] 1.644 $\mu$m line is quite strong and isolated, we still considered the effects of neighboring features. The red wing of the NIR [Fe II] 1.644 $\mu$m feature may be affected by [Fe II] 1.745 $\mu$m and [Co III] 1.741 and 1.764 $\mu$m. Maguire et al. (2018) choose to use the blended [Fe II] feature near $\sim$1.2 $\mu$m instead of the [Fe II] 1.644 $\mu$m feature based on concerns of contamination from these Co lines.

But at such late phases, Co lines are expected to be drastically diminished as Co decays, leaving [Fe II] 1.745 $\mu$m as the main contaminant. The peak of the feature near 1.54 $\mu$m was also observed to shift toward the blue, presumably from being [Co II] 1.547 $\mu$m and [Co III] 1.549 $\mu$m dominated to being [Fe II] 1.534 $\mu$m dominated, indicating a diminishment of these Co features. The time-series spectroscopy presented in this work also extends to phases much later than the data set used by Maguire et al. (2018), which allows more time for Co decay and further minimizes concerns of contamination from Co lines in the 1.64 $\mu$m region.

Since Co lines are significantly weaker by these late phases, the [Fe II] 1.745 $\mu$m is likely the main contaminant while fitting the [Fe II] 1.644 $\mu$m feature. Note that this only affects the red wing of the feature and has been examined by testing the effects of different line profile and continuum functions (Figure 6). Variations in line profile function, continuum, and the choice of fit region boundaries can affect the resulting line-center measurements more than contamination from the [Fe II] 1.745 $\mu$m line.

Figure 6 shows the differences between different continuum and line profile combinations used to fit the [Fe II] 1.644 $\mu$m feature in the NIR. The reduced $\chi^2$ per degree of freedom has been used to quantify the robustness of fitting methods throughout this work and is noted next to each fit result. These reduced $\chi^2$ values show that optical fitting methods are more susceptible to underfitting the data (see Figures 3 and 4). The NIR does not underfit the data for any combination of continuum and line profile type. Therefore, the NIR provides more consistent velocity measurements regardless of fitting method. The combination of reduced $\chi^2$ values and smaller variation in velocity measurement results leads us to conclude that the NIR provides a more robust measurement of [Fe II] line velocities than the optical.

To summarize, when compared to the optical, the NIR [Fe II] velocity measurements are robust against the following:

1. variations in the boundary, continuum, and profile function selections for the profile fits;
2. fit degeneracy due to line blending; and
3. emerging spectral features.

### 4.2. Explosion Kinematics

There are relatively few NIR spectra published at such late phases, but there are other studies that have found similarly low 1.644 $\mu$m line velocities. Diamond et al. (2018) measure an [Fe II] 1.644 $\mu$m line velocity of 1330 km s$^{-1}$ for SN 2014J at +370 days. An earlier study published a nebular-phase spectrum of SN 2005df at +380 days and reports an [Fe II] 1.644 $\mu$m line velocity shift of $\sim$530 km s$^{-1}$ (Diamond et al. 2015). Furthermore, Diamond et al. (2015) determine an upper limit for the [Fe II] 1.644 $\mu$m line velocity of SN 2005df to be no greater than 1200 km s$^{-1}$.

Our resulting velocity measurements for SN 2013aa show the same velocity shift directions as those measured by Maguire et al. (2018), which used the blended set of [Fe II] lines near $\sim$1.2 $\mu$m instead of the [Fe II] 1.644 $\mu$m line. However, the NIR [Fe II] velocity shifts measured using the $\sim$1.2 $\mu$m feature are consistently larger than those from the 1.644 $\mu$m feature, despite being measured from the same observed spectrum. This may be due to line blending and multi-Gaussian fit degeneracies that can affect measurement results (Graham et al. 2022).

For SN 2013aa, Maguire et al. (2018) report [Fe II] $\lambda$7155 line velocities of $-771 \pm 206$ km s$^{-1}$ at +360 days and $-385 \pm 244$ km s$^{-1}$ at +425 days, as well as [Fe II] 1.2567 $\mu$m velocities of $-1408 \pm 323$ km s$^{-1}$ at +360 days and $-1189 \pm 637$ km s$^{-1}$ at +425 days.

It is interesting to note that the optical [Fe II] $\lambda$7155 velocities reported by Maguire et al. (2018) are in better agreement with our [Fe II] 1.644 $\mu$m velocities (see Figure 7). We emphasize that these three [Fe II] velocity measurements were all obtained using the same XShooter spectrum. This demonstrates how different choices made during the fitting process can yield results with possibly contradictory science implications.

All the published NIR [Fe II] velocity measurements appear to be much lower than the [Fe II] velocities measured using optical features, which are typically thousands of kilometers per second (e.g., Maeda et al. 2010b; Graham et al. 2017). One possible explanation for the discrepant velocity measurements may be that the NIR is probing deeper into the central region than the optical. If this is true, it again makes the NIR a more accurate probe of the explosion kinematics.

One of the main results of this work is the low Doppler shift velocity measured from the NIR. Since we are measuring the velocity of the material in the innermost regions of the SNe Ia, these consistently low velocities could be used to rule out certain explosion scenarios. For example, in the D$^6$ scenario (Shen et al. 2018b), the surviving companion WD and the SN remnant are expected to have relatively high velocities of at least 1000 km s$^{-1}$. The low velocities of SN 2013aa and SN 2017cbv are not consistent with this, unless both systems are oriented face-on. Furthermore, the nebular-phase spectra do not show multiple double-peaked emission features that are the possible signature of a head-on collision (e.g., Dong et al. 2015). This was most clearly shown in the NIR, as a single-Gaussian profile provided good fits.

### 4.3. Viewing Angle

Both SN 2013aa and SN 2017cbv are classified as LVG objects (Graham et al. 2017, 2022), based on the time evolution of the Si II $\lambda$6355 velocity as defined by Maeda et al. (2010a). In the context of their off-center DDT model, HVG SNe Ia are mainly associated with redshifted nebular-phase emission lines, whereas LVG SNe Ia can exhibit a range of velocity shifts.

Based on the optical spectra alone, SN 2013aa exhibits a sizable blueshift in the nebular-phase [Fe II] lines, while SN 2017cbv shows a smaller redshift. Due to the large uncertainties in the optical measurements, the direction of the shift for SN 2017cbv is not conclusive based on optical spectra alone. Nonetheless, these results are largely consistent with the off-center DDT predictions laid out above.

The NIR measurements generally yield the same nebular velocity shift directions as the optical. However, the absolute values of the velocity shifts are significantly lower. In the case of SN 2013aa, the NIR velocity shift progressively decreases with time until it is consistent with zero past +500 days. This may point to a geometric dilution effect, while the last NIR spectrum points to no significant offset of the initial ignition, in contrast to the $\sim$1000 km s$^{-1}$ shift given by the optical spectra. Recall the limitations for measuring the velocity shifts using the optical region outlined in Section 4.1. The trend in NIR shifts with time is not as clear in SN 2017cbv, due to the larger velocity measurement error.

Viewing angle effects can be further investigated in the NIR. By taking advantage of the isolated [Fe II] 1.644 $\mu$m line, it is also possible to examine the geometry of the inner region using the shape of the line profile (e.g., Diamond et al. 2018; Hoeflich et al. 2021). In the most extreme case of an off-center ignition, this line profile can appear tilted and highly asymmetric, with a significantly shifted peak (Hoeflich et al. 2021). As noted previously, we do not detect significant asymmetries in the [Fe II] 1.644 $\mu$m line in SN 2013aa and SN 2017cbv.

If the Doppler shifts of the nebular lines for both SNe Ia are nonzero, both the optical and NIR measurements suggest that the shifts for the sibling SNe Ia are in opposite directions. Given the remarkable similarities between SN 2013aa and SN 2017cbv at early times (Burns et al. 2020), it is likely that the shifts indeed come from viewing angle effects. With the higher confidence in the NIR measurements, we suggest that the offsets in the ignition points are small in these two SNe Ia. This result highlights the importance of increasing the NIR nebular sample to assess whether these small shifts are common in SNe Ia.

### 4.4. Ionization State

A time-series data set extending beyond +500 days allows for the examination of the ionization state in the central region of the SN. The most prominent optical emission feature in the nebular phase is near 4700 Å and is mainly attributed to [Fe III]. It is commonly detected through very late phases for most SNe Ia and is one of the main optical identifiers signifying the nebular phase. This feature can be used in relation to other [Fe II] and [Fe III] features in both the optical and the NIR to gauge the ionization state.

The emission feature near 4700 Å is mainly composed of three [Fe III] lines and one [Fe II] line: [Fe III] $\lambda\lambda$4658, 4701, and 4734, as well as [Fe II] $\lambda$4814, with the strongest contribution from [Fe III] (e.g., Mazzali et al. 2011, 2015). This feature is often compared to the emission feature near 5200 Å, which is mainly composed of three [Fe II] lines and one [Fe III] line: [Fe II] $\lambda\lambda$5159, 5262, and 5333, as well as [Fe III] $\lambda$5270 (e.g., Mazzali et al. 2011, 2015). Changes in the relative strengths of the $\lambda\lambda 4700$ and 5200 features could indicate a change in the ionization state.

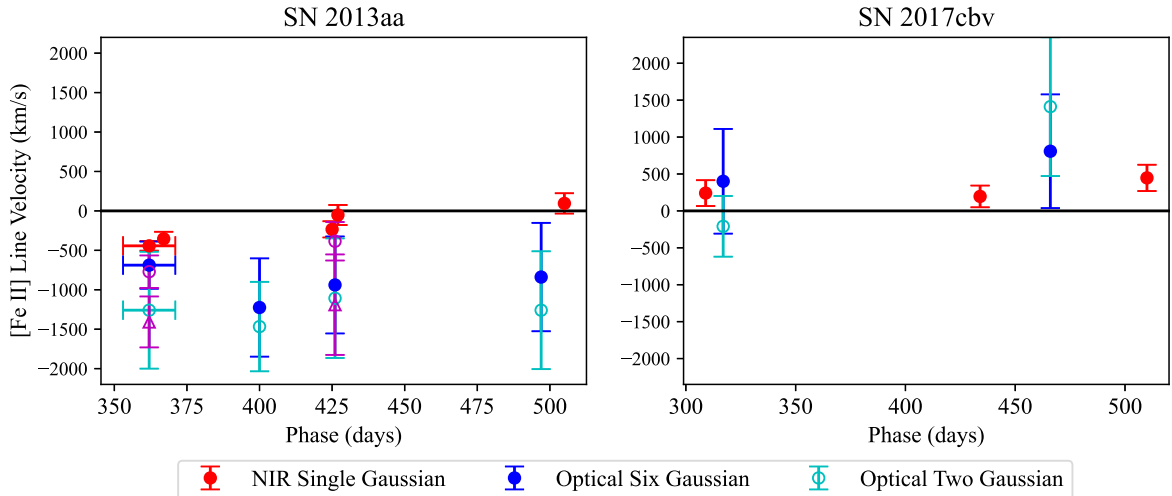

**Figure 7.** The line-center velocity of [Fe II] measured using [Fe II] $\lambda7155$ in the optical and [Fe II] 1.644 $\mu$m in the NIR. NIR measurements are shown in red. Optical measurements are shown with cyan open and blue filled circles for two- and six-Gaussian fits, respectively. NIR measurements are far more robust and show systematically lower velocities in the absolute sense than the optical measurements. For comparison, we include [Fe II] line velocity measurements of SN 2013aa from Maguire et al. (2018), which were obtained using the same XShooter spectra used in this study. The magenta markers show results of their multi-Gaussian fits for the optical [Fe II] $\lambda7155$ line (open magenta circles) and NIR [Fe II] 1.2567 $\mu$m line (open magenta triangles). Although the directions of the velocity shifts are in agreement, the [Fe II] $\lambda7155$ and [Fe II] 1.2567 $\mu$m line velocities are consistently larger than the velocities measured using the [Fe II] 1.644 $\mu$m line.

In the earliest optical spectra in our sample the $\lambda 4700$ feature is clearly detected in both SNe. As the feature evolves in time, the strength appears to weaken, until it significantly diminishes by +497 days in SN 2013aa. Note, however, that the +497-day spectrum of SN 2013aa has more noise than the other optical spectra in our sample. The $\lambda 4700$ feature is prominent in our first optical spectrum of SN 2017cbv at +317 days, but by +466 days the $\lambda 4700$ feature is drastically diminished but still present, while the $\lambda 5200$ feature stays strong and clearly detectable. SN 2017cbv at +466 days also shows increased emission on the blue side of the $\lambda 4700$ feature, which may be emission from nearby [Fe II] lines: $\lambda\lambda 4224$ and 4416 (Mazzali et al. 2011, 2015). From the optical spectra alone, the time evolution of the $\lambda\lambda 4700$ and 5200 features of both SNe Ia indicates a change in the ionization state by +500 days.

NIR spectra can be used to further investigate this time evolution observed in the optical. NIR features can potentially provide a cleaner separation between contributions from [Fe II] and [Fe III]. We again make use of the strong and isolated [Fe II] 1.644 $\mu$m line. For [Fe III], the emission feature near 2.2 $\mu$m is a good candidate and is primarily composed of four [Fe III] lines: 2.1457, 2.2184, 2.2427, and 2.3485 $\mu$m (Diamond et al. 2018). However, this feature is often not well observed due to the lower throughput and thermal background in the K band.

This NIR [Fe III] feature is detected in our earliest NIR nebular-phase spectra for both SNe. By the second NIR observation (+428 days and +434 days for SN 2013aa and SN 2017cbv, respectively), this feature has clearly weakened and is no longer detectable within the noise. The optical and NIR results are consistent, pointing to an evolving ionization state with notable changes in spectral features occurring between +400 and +500 days.

Previous studies also showed this shift in the ionization state (e.g., Mazzali et al. 2020; Tucker et al. 2022). In SN 2011fe, the prominent optical [Fe III] feature near 4700 Å disappeared completely by +576 days (Figure 8; Taubenberger et al. 2015). Using late-time luminosities of SN 2011fe from Kerzendorf et al. (2014) and nebular-phase optical spectroscopy, Taubenberger et al. (2015) concluded that thermal excitation alone is not enough to reproduce the observed emission features. Thus, the disappearance of [Fe III] in SN 2011fe is likely due to recombination or other nonthermal excitation processes. For SN 2014J at +435 days, there is a drastically diminished $\lambda 4700$ feature compared to earlier phases, with the $\lambda 5200$ feature remaining strong (Figure 8; Mazzali et al. 2020). On the other hand, the 2.2 $\mu$m feature in SN 2014J remains relatively undiminished when compared to SN 2013aa and SN 2017cbv.

The time evolution of the peak flux ratios in the optical and NIR is presented in Figure 9. The flux ratios were calculated using the $\lambda\lambda 4700$ and 5200 features in the optical and the 2.2 $\mu$m feature and the 1.644 $\mu$m line in the NIR. The relative strengths of [Fe III] and [Fe II] of SN 2013aa and SN 2017cbv, as indicated by the flux ratios, have very similar time evolution in both the optical and the NIR. In the optical, SN 2013aa and SN 2017cbv share strong similarities with SN 2011fe. SN 2014J appears to be the outlier in this small sample, with the optical indicating a dearth of [Fe III] and the NIR indicating the opposite. The NIR, with less severe line blending than the optical, can potentially provide a more robust result.

### 4.5. Central Density and Magnetic Field

Since the [Fe II] 1.644 $\mu$m line is strong and well isolated, it has been used to study the subtle effects of the initial magnetic field and central density of the progenitor WD. The central density of the WD determines the amount of electron capture within the central region where stable iron-group elements (IGEs) are produced during the deflagration phase of the thermonuclear runaway. The stable IGEs produced in this region create a radioactive "hole" in the ejecta. The width and shape of the [Fe II] 1.644 $\mu$m line are determined by the size and geometry of this hole. Diamond et al. (2015) modeled a range of central densities and showed that a higher central density would produce a broader [Fe II] 1.644 $\mu$m profile, since the radioactive $^{56}$Ni would be pushed outward to higher velocities, thus broadening the emission line (Penney & Hoeflich 2014).

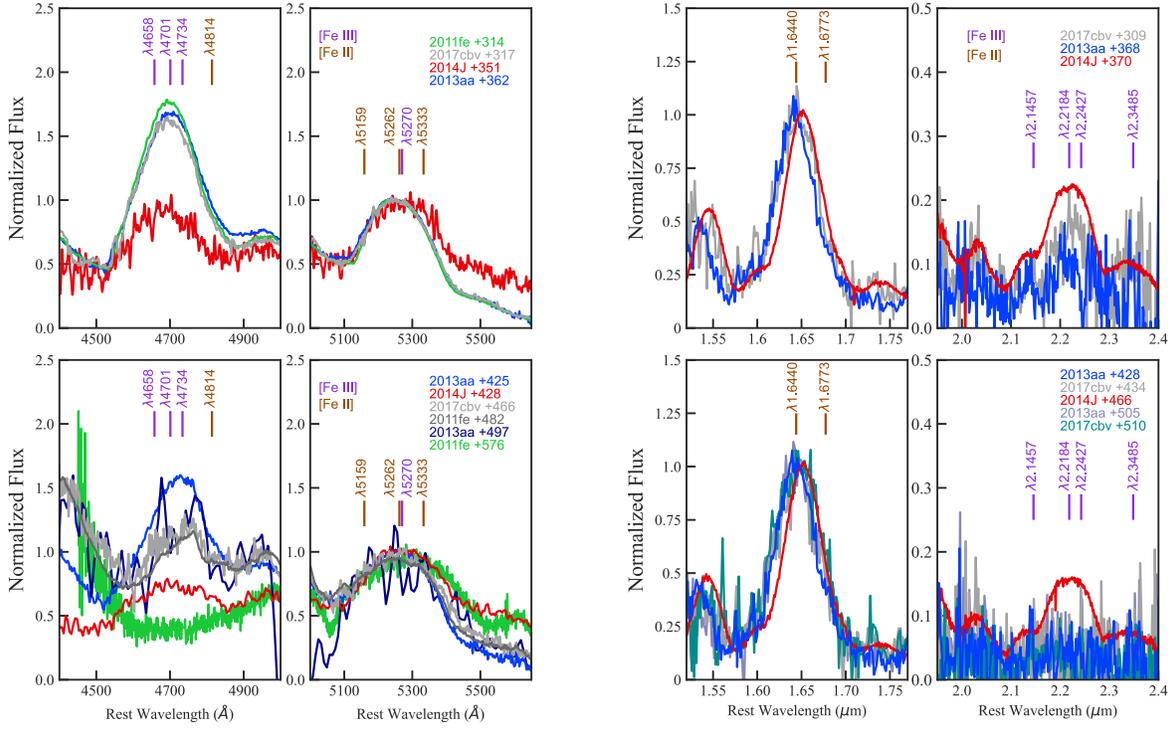

**Figure 8.** Time evolution of [Fe II] and [Fe III] features in the optical and NIR, with spectra before +400 days (top panels) and after +400 days (bottom panels). For comparison, we use previously published optical spectra of SN 2011fe (Graham et al. 2015; Taubenberger et al. 2015; Tucker et al. 2022) and SN 2014J (Srivastav et al. 2016; Zhang et al. 2018) and NIR spectra of SN 2014J (Diamond et al. 2018). The spectra are normalized to the peak of the features dominated by [Fe II] (the λ5200 feature in the optical and the [Fe II] 1.644 μm line in the NIR). Note that SN 2014J also exhibits a persistent redshifted [Fe II] 1.644 μm line as reported by Diamond et al. (2018).

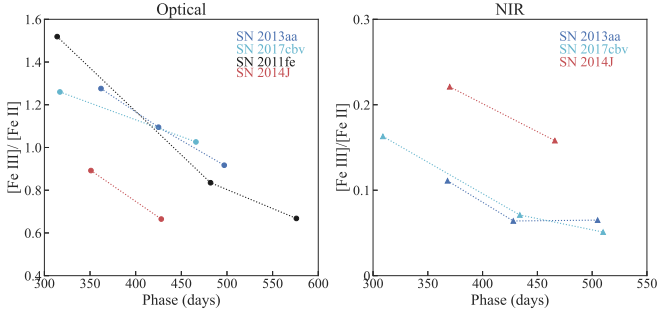

**Figure 9.** Time evolution of the [Fe III]-to-[Fe II] flux ratios of prominent emission features. The left panel shows the optical flux ratios of the λ4700 feature (dominated by [Fe III]) and the λ5200 feature (dominated by [Fe II]). The right panel shows the NIR flux ratios of the 2.2 μm feature (dominated by [Fe III]) and the [Fe II] 1.644 μm line. Both show a steady decline of [Fe III] strength, demonstrating how the optical and NIR can be used to corroborate each other.

The time evolution of the [Fe II] 1.644 μm line can also indicate the presence and strength of the progenitor magnetic field. By +300 days, the dominant source of energy transport within the SN Ia are positrons from the beta decay of $^{56}$Co (Höflich et al. 2004; Penney & Hoeflich 2014). In the presence of a strong magnetic field, the positrons would stay trapped in the central region. The [Fe II] 1.644 μm line would then stay broad until the magnetic field weakens after approximately +500 days after peak brightness and the positrons can escape, resulting in a narrower line profile. If the [Fe II] 1.644 μm line becomes narrow well before +500 days, it could indicate a weaker magnetic field. The presence of a magnetic field is expected to influence the thermonuclear runaway and, thus, the smoldering phase, deflagration, turbulence, and detonation.

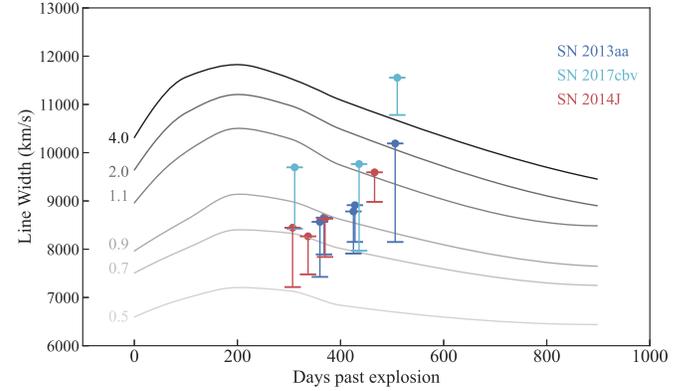

**Figure 10.** Time evolution of the line width of [Fe II] 1.644 μm feature. The gray curves correspond to the predicted line width evolution from non-LTE DDT models of Diamond et al. (2018). The gray colors correspond to different WD central densities that span a range of (0.5–4.0) × 10$^9$ g cm$^{-3}$. Measured line widths of SN 2013aa, SN 2017cbv (this work), and SN 2014J (Diamond et al. 2018) are shown for comparison. The error bars represent the extreme line width values based on different continuum subtractions and do not represent the typical 1σ uncertainty. All three SNe Ia show sustained and even increasing width, indicating high central densities or possibly magnetic fields.

In Figure 10, our measured line widths are shown along with the line width evolution predicted by DDT models with a range of central densities (Diamond et al. 2018). For clarity, we display only a subset of these DDT models that have been averaged over all viewing angles. The error bars represent the extreme line width values based on different continuum subtractions and do not represent the typical $1\sigma$ uncertainty.

Based on these models, we estimate that the two SNe Ia have central densities no lower than $0.5 \times 10^9$ g cm$^{-3}$ and no higher than $4.0 \times 10^9$ g cm$^{-3}$. SN 2017cbv has a consistently broader [Fe II] line than SN 2013aa, which may indicate that the WD progenitor of SN 2017cbv had a slightly higher central density, although the uncertainties due to continuum subtraction are large.

The [Fe II] 1.644 $\mu$m feature becomes wider between +300 and +500 days for both SN 2013aa and SN 2017cbv. This sustained widening continues much later than any of the DDT models shown here and may indicate strong initial magnetic fields in both SNe Ia. Stronger magnetic fields may help prolong the widening in the models (Penney & Hoeflich 2014) to match the observed behavior.

In comparison to the previously published measurements for SN 2014J, the time evolution of the line width is quite similar. The [Fe II] 1.644 $\mu$m feature of SN 2014J also continues to broaden beyond +400 days (Figure 9 in Diamond et al. 2018). A similar evolution is observed in three different well-observed SNe Ia using two different instruments and at multiple phases. This indicates that the effect is real and not an observational artifact. The minimum possible line widths also indicate that all three of these SNe Ia are unlikely to have very low progenitor WD central densities and masses.

## 5. Conclusion

We present nebular-phase time-series spectroscopy of the siblings SN 2013aa and SN 2017cbv, two SNe Ia located in the same host galaxy. The SNe Ia are located in the outskirts of a nearby, well-studied host galaxy, NGC 5643, which allowed the direct comparison of their properties without some of the uncertainties that may arise when comparing two SNe located in two different host galaxies. The spiral host appears nearly face-on, yet the velocity contribution of its rotation is large enough to affect nebular-phase line velocity measurements, especially in the NIR.

During the nebular phase, the most isolated strong forbidden emission feature is the [Fe II] 1.644 $\mu$m line. The time evolution of the line width can be used to investigate progenitor properties, such as the initial central density and magnetic field. The two SNe Ia are very similar at early times, and most of their similarities persist through the nebular phase, with a few notable differences. For example, the line width measurements suggest a slight difference in the central density, though both are high enough to form stable IGEs in the innermost regions. The sustained broadening of the [Fe II] 1.644 $\mu$m line also implies the possibility of high magnetic fields in the progenitors of both SNe Ia.

Our multiwavelength data set is ideal for the comparisons of optical and NIR explosion kinematics measurements. Several fitting methods, commonly adopted to find the centroids and Doppler shifts of the [Fe II] features, were tested and compared. In the optical, the most widely used feature is the one dominated by [Fe II] $\lambda$7155. This feature is heavily blended by other [Fe II] and [Ni II] lines. Thus, a common solution is to fit the feature with multiple components. We found that by adopting a multicomponent fit, the optical measurements are susceptible to variations in the selection of the fit boundary and profile functions. Furthermore, an emerging [Ca II] line near the center of this feature is clearly evident in one of our spectra and further degrades the accuracy. On the other hand, the relatively isolated NIR [Fe II] 1.644 $\mu$m line is free of these issues and was shown to be more robust and accurate.

Most significantly, the NIR measurements consistently yield substantially slower velocities than the optical ones. For both SN 2013aa and SN 2017cbv, the NIR measurements yield Doppler shifts less than 500 km s$^{-1}$, much smaller than the $\sim$1000 km s$^{-1}$ the optical measurements suggest. The discrepancy points to a potential systematic for studies that rely solely on the optical region. It could also indicate that the NIR is probing a deeper layer, in which case the NIR is again preferred over the optical for observing the innermost regions of the SN. Unless both progenitor systems are oriented nearly face-on to us, the low velocities can rule out most close double-degenerate scenarios, such as D$^6$ and head-on collisions, which would result in high velocities for both the companion and the SN remnant. Our results here highlight the need for a larger NIR sample to determine whether these properties are unique to these two SNe Ia.

We thank the Las Campanas technical staff for their continued support over the years. The CSP-II has been supported by NSF grants AST-1008343, AST-1613426, AST-1613455, and AST-1613472, as well as the Danish Agency for Science and Technology and Innovation through a Sapere Aude Level 2 grant.

This research has made use of the services of the ESO Science Archive Facility, based on observations collected at the European Southern Observatory under ESO program 095.B-0532. The Australia Telescope Compact Array is part of the Australia Telescope National Facility, which is funded by the Australian Government for operation as a National Facility managed by CSIRO. We acknowledge the Gomeroi people as the traditional owners of the Observatory site.

S.K. was supported by the Florida Space Consortium Research Grant. L.G. acknowledges financial support from the Spanish Ministerio de Ciencia e Innovación (MCIN), the Agencia Estatal de Investigación (AEI) 10.13039/501100011033, and the European Social Fund (ESF) "Investing in your future" under the 2019 Ramón y Cajal program RYC2019-027683-I and the PID2020-115253GA-I00 HOSTFLOWS project, from Centro Superior de Investigaciones Científicas (CSIC) under the PIE project 20215AT016, and the program Unidad de Excelencia María de Maeztu CEX2020-001058-M. E.B. was supported in part by NASA grant 80NSSC20K0538. P.H. is supported by the NSF grant AST-1715133. This research has made use of the NASA/IPAC Extragalactic Database (NED), which is funded by the National Aeronautics and Space Administration and operated by the California Institute of Technology.

*Facilities:* Magellan:Baade (FIRE), Gemini South (GMOS), VLT (XShooter), ATCA (375 m and 1.5 A arrays), Magellan: Clay (LDSS-3).

*Software:* astropy (Collaboration et al. 2013; Astropy Collaboration et al. 2018), LMFIT (Newville et al. 2014).

### ORCID iDs

Sahana Kumar ⓘ https://orcid.org/0000-0001-8367-7591
Eric Y. Hsiao ⓘ https://orcid.org/0000-0003-1039-2928
C. Ashall ⓘ https://orcid.org/0000-0002-5221-7557
M. M. Phillips ⓘ https://orcid.org/0000-0003-2734-0796
N. Morrell ⓘ https://orcid.org/0000-0003-2535-3091

P. Hoeflich https://orcid.org/0000-0002-4338-6586
C. R. Burns https://orcid.org/0000-0003-4625-6629
L. Galbany https://orcid.org/0000-0002-1296-6887
E. Baron https://orcid.org/0000-0001-5393-1608
C. Contreras https://orcid.org/0000-0001-6293-9062
S. Davis https://orcid.org/0000-0002-2806-5821
F. Förster https://orcid.org/0000-0003-3459-2270
M. L. Graham https://orcid.org/0000-0002-9154-3136
E. Karamehmetoglu https://orcid.org/0000-0001-6209-838X
R. P. Kirshner https://orcid.org/0000-0002-1966-3942
B. Koribalski https://orcid.org/0000-0003-4351-993X
K. Krisciunas https://orcid.org/0000-0002-6650-694X
J. Lu https://orcid.org/0000-0002-3900-1452
P. J. Pessi https://orcid.org/0000-0002-8041-8559
A. L. Piro https://orcid.org/0000-0001-6806-0673
M. Shahbandeh https://orcid.org/0000-0002-9301-5302
M. D. Stritzinger https://orcid.org/0000-0002-5571-1833
N. B. Suntzeff https://orcid.org/0000-0002-8102-181X
S. A. Uddin https://orcid.org/0000-0002-9413-4186